\documentclass[a4paper,12pt]{article}

\pdfoutput=1
\usepackage{amsmath}
\usepackage{amssymb}
\usepackage{amsfonts}
\usepackage{mathrsfs}
\usepackage{bbm}
\usepackage{graphicx,subfigure,booktabs}
\usepackage[numbers,sort&compress]{natbib}
\usepackage{verbatim}
\usepackage{color}
\usepackage{ulem}
\usepackage{setspace}
\usepackage{multirow}
\usepackage{url}
\usepackage{enumerate}

\usepackage[utf8]{inputenc}

\usepackage[colorlinks,
linkcolor=black,
filecolor=black,
anchorcolor=black,
urlcolor=black,
citecolor=blue,
bookmarks=false,
]{hyperref}

\usepackage[hyphenbreaks]{breakurl}

\numberwithin{equation}{section}

\newcommand{\kkb}{$K^0-\bar{K}^0$}
\newcommand{\cp}{$CP$}

\newlength{\dinwidth}
\newlength{\dinmargin}
\makeatletter
\newcommand{\thickhline}{%
	\noalign {\ifnum 0=`}\fi \hrule height 1pt
	\futurelet \reserved@a \@xhline
}
\makeatother
\allowdisplaybreaks

\begin{document}

\title{\bf \boldmath $CP$ asymmetry in the angular distribution of $\tau\to K_S\pi\nu_\tau$ decays}

\author{Feng-Zhi Chen\footnote{ggchan@mails.ccnu.edu.cn},
        Xin-Qiang Li\footnote{Corresponding author: xqli@mail.ccnu.edu.cn},
    and Ya-Dong Yang\footnote{yangyd@mail.ccnu.edu.cn}\\[12pt]
\small Institute of Particle Physics and Key Laboratory of Quark and Lepton Physics~(MOE), \\
\small Central China Normal University, Wuhan, Hubei 430079, P.~R. China}

\date{}
\maketitle
\vspace{-0.2cm}

\begin{abstract}
\noindent In this work, we study the $CP$ asymmetry in the angular distribution of $\tau\to K_S\pi\nu_\tau$ decays, taking into account the known $CP$ violation in $K^0-\bar{K}^0$ mixing. It is pointed out for the first time that, once the well-measured $CP$ violation in the neutral kaon system is invoked, a non-zero $CP$ asymmetry would appear in the angular observable of the decays considered, even within the Standard Model. By employing the reciprocal basis, which is most convenient when a $K_{S(L)}$ is involved in the final state, the $CP$-violating angular observable is derived to be two times the product of the time-dependent $CP$ asymmetry in $K\to \pi^+\pi^-$ and the mean value of the angular distribution in $\tau^\pm\to K^0(\bar{K}^0)\pi^\pm\bar{\nu}_\tau(\nu_\tau)$ decays. Compared with the Belle results measured in four different bins of the $K\pi$ invariant mass, our predictions lie within the margins of these measurements, except for a $1.7~\sigma$ deviation for the lowest mass bin. While being below the current Belle detection sensitivity that is of $\mathcal{O}(10^{-3})$, our predictions are expected to be detectable at the Belle II experiment, where $\sqrt{70}$ times more sensitive results will be obtained with a $50~\text{ab}^{-1}$ data sample.
\end{abstract}

\newpage

\section{Introduction}
\label{sec:intro}
\cp\ violation in weak interactions has now been observed in the $K$-, $B$-, and $D$-meson systems~\cite{Christenson:1964fg,AlaviHarati:1999xp,Lai:2001ki,Aubert:2001nu,Abe:2001xe,Aubert:2004qm,Chao:2004mn,Aaij:2012kz,Aaij:2013iua,Aaij:2019kcg}, and all results could be accommodated within the Standard Model (SM) by the single irreducible complex phase in the Cabibbo-Kobayashi-Maskawa (CKM) quark-mixing matrix~\cite{Cabibbo:1963yz,Kobayashi:1973fv}. However, the fundamental origin of \cp\ violation is still unknown, and it is important to look for other $CP$-violating effects in as many systems as possible. One such system is the $\tau$ lepton, which is the only known lepton massive enough to decay into hadrons. In this regard, the hadronic $\tau$ decays, besides serving as a clean laboratory for studying various low-energy aspects of the strong interaction~\cite{Pich:2013lsa,Davier:2005xq}, may also allow us to explore $CP$-violating effects both within and beyond the SM~\cite{Tsai:1996ps,Bigi:2012km,Bigi:2012kz,Kiers:2012fy}.    

Searches for \cp\ violation in hadronic $\tau$ decays have been carried out by several experiments. After the initial null results from CLEO~\cite{Anderson:1998ke,Bonvicini:2001xz} and Belle~\cite{Bischofberger:2011pw} in the search for \cp\ violation in the angular distribution of $\tau\to K_S\pi\nu_\tau$ decays, a non-zero \cp\ asymmetry was reported for the first time by the BaBar collaboration~\cite{BABAR:2011aa}, by measuring the decay-rate difference between $\tau^+\to K_S\pi^+\bar\nu_\tau$ and $\tau^-\to K_S\pi^-\nu_\tau$ decays. Within the SM, this asymmetry is predicted to be non-zero~\cite{Bigi:2005ts,Calderon:2007rg,Grossman:2011zk}, due to the \cp\ violation in \kkb\ mixing~\cite{Christenson:1964fg,Tanabashi:2018oca}, but is found to be $2.8~\sigma$ away from the BaBar measurement~\cite{BABAR:2011aa}. Such a discrepancy has motivated many studies of possible direct \cp\ asymmetries in $\tau\to K_S\pi\nu_\tau$ decays induced by non-standard tensor interactions~\cite{Devi:2013gya,Dhargyal:2016kwp,Dhargyal:2016jgo,Dighe:2019odu,Cirigliano:2017tqn,Rendon:2019awg,Chen:2019vbr}, finding that such a large asymmetry cannot be explained by a tensor coupling under the combined constraints from other observables~\cite{Cirigliano:2017tqn,Rendon:2019awg,Chen:2019vbr}. With the integrated luminosity of the data sample increased, it is expected that measurements of $CP$-violating effects in hadronic $\tau$ decays will be improved at the Belle II experiment~\cite{Kou:2018nap}. This in turn motivates us to improve further the corresponding theoretical predictions.

In this work, we shall focus on the \cp\ asymmetry in the angular distribution of $\tau\to K_S\pi\nu_\tau$ decays, which cannot be observed from measurements of the $\tau^{\pm}$ decay rates, but is detected as a difference in the $\tau^{\pm}$ decay angular distributions, without requiring information about the $\tau$ polarization or the determination of the $\tau$ rest frame~\cite{Tsai:1996ps,Kuhn:1990ad,Kuhn:1996dv,Kimura:2009pm,Kimura:2014wsa}. Following the notations adopted by the Belle collaboration~\cite{Bischofberger:2011pw}, we can express this $CP$-violating observable as 
\begin{align}\label{eq:ACP_i}
A_i^{CP}=\frac{\int_{s_{1,i}}^{s_{2,i}}\int_{-1}^{1} \cos \alpha \left[\frac{d^2 \Gamma(\tau^-\to K_S\pi^-\nu_\tau)}{ds\,d\cos\alpha} -\frac{d^2 \Gamma(\tau^+\to K_S\pi^+\bar{\nu}_\tau)}{ds\,d\cos\alpha}\right] ds\,d\cos\alpha}{\frac{1}{2}\int_{s_{1,i}}^{s_{2,i}}\int_{-1}^{1}\left[\frac{d^2 \Gamma(\tau^-\to K_S\pi^-\nu_\tau)}{ds\,d\cos\alpha}+\frac{d^2 \Gamma(\tau^+\to K_S\pi^+\bar{\nu}_\tau)}{ds\,d\cos\alpha}\right] ds\,d\cos\alpha}\,,
\end{align}
which is defined as the difference of the differential $\tau^-$ and $\tau^+$ decay widths weighted by $\cos\alpha$, and can be evaluated in bins of the $K\pi$ invariant mass squared $s$, with the $i$-th bin defined by the interval $[s_{1,i},s_{2,i}]$~\cite{Bischofberger:2011pw}. Here $\alpha$ is the angle between the directions of $K$ and $\tau$ as seen in the $K\pi$ rest frame~\cite{Bischofberger:2011pw,Tsai:1996ps,Kuhn:1996dv}. When considering the observable $A_i^{CP}$, one should keep in mind the following two facts~\cite{Grossman:2011zk}: (i) the $\tau^+$~($\tau^-$) decay produces initially a $K^0$~($\bar{K}^0$) state due to the $\Delta S=\Delta Q$ rule; (ii) the intermediate state $K_S$ is not observed directly in experiment, but rather reconstructed in terms of a $\pi^+\pi^-$ final state with $M_{\pi\pi}\approx M_K$ and a time that is close to the $K_S$ lifetime. However, since \cp\ is violated in \kkb\ mixing, the final state $\pi^+\pi^-$ can be obtained not only from the short-lived $K_S$ but also from the long-lived $K_L$ state. Thus, the measured \cp\ asymmetry depends sensitively on the kaon decay time interval over which it is integrated. As emphasized by Grossman and Nir~\cite{Grossman:2011zk} in the study of the decay-rate asymmetry, the contribution from the interference between the amplitudes of intermediate $K_S$ and $K_L$ is not negligible, but is as important as the pure $K_S$ amplitude. In addition, one should also take into account the experiment-dependent effects, due to the efficiency as a function of the kaon decay time as well as the kaon energy in the laboratory frame to account for the time dilation~\cite{Grossman:2011zk}.

It is known that, after neglecting the effect generated by the second-order weak interaction, which is estimated to be of $\mathcal{O}(10^{-12})$~\cite{Delepine:2005tw}, there exists direct \cp\ violation neither in the decay rate nor in the angular distribution of $\tau^\pm\to K^0(\bar{K}^0)\pi^\pm\bar{\nu}_\tau(\nu_\tau)$ decays within the SM~\cite{Tsai:1996ps,Kuhn:1996dv}. Once the known \cp\ violation in the neutral kaon system is taken into account, however, a non-zero \cp\ asymmetry is predicted in the decay rates~\cite{Bigi:2005ts,Calderon:2007rg,Grossman:2011zk}. Here we shall investigate for the first time whether an observable \cp\ asymmetry in the angular distribution of $\tau\to K_S\pi\nu_\tau$ decays could be generated by the well-measured \cp\ violation in \kkb mixing. To this end, we shall employ the reciprocal basis~\cite{Sachs:1963zz,Enz:1965tr,Wolfenstein:1970wb,Branco:1999fs,Silva:2000db,Silva:2004gz}, which is most convenient when a $K_{S(L)}$ is involved in the final state and has been used to reproduce conveniently the decay-rate asymmetry~\cite{Chen:2019vbr}. It is then found that a non-zero $CP$ asymmetry would appear in the angular observable of the decays considered, even within the SM. Furthermore, this observable is derived to be two times the product of the time-dependent $CP$ asymmetry in $K\to \pi^+\pi^-$ and the mean value of the angular distribution in $\tau^\pm\to K^0(\bar{K}^0)\pi^\pm\bar{\nu}_\tau(\nu_\tau)$ decays. While the formalism of the former is quite clear~\cite{Branco:1999fs}, a precise description of the latter requires information about the $K\pi$ vector and scalar form factors, including both their moduli and phases. As the form-factor phases fitted via a superposition of Breit-Wigner functions with complex coefficients do not vanish at threshold and violate Watson's final-state interaction theorem before the higher resonances start to play an effect~\cite{Cirigliano:2017tqn,Rendon:2019awg,Chen:2019vbr}, we cannot rely on the formalism developed in Refs.~\cite{Epifanov:2007rf,Paramesvaran:2009ec,Finkemeier:1995sr,Finkemeier:1996dh} to study the $CP$ asymmetry in $\tau\to K_S\pi\nu_\tau$ decays. Instead, we shall adopt the thrice-subtracted (for the vector form factor)~\cite{Boito:2008fq,Boito:2010me} and the coupled-channel (for the scalar form factor)~\cite{Jamin:2000wn,Jamin:2001zq,Jamin:2006tj} dispersive representations, which warrant the properties of unitarity and analyticity, and contain a full knowledge of QCD in both the perturbative and non-perturbative regimes.

With all the above points taken into account, we shall then present our predictions for the $CP$-violating angular observable $A_i^{CP}(t_1,t_2)$ defined by Eq.~\eqref{eq:reduceACPi}. It should be emphasized again that our presentation is confined to the SM framework. This is totally different from the studies made in Refs.~\cite{Anderson:1998ke,Bonvicini:2001xz,Bischofberger:2011pw}, which are aimed to probe possible $CP$-violating (pseudo-)scalar couplings beyond the SM~\cite{Tsai:1996ps,Kuhn:1996dv,Kimura:2009pm,Kimura:2014wsa,Kiers:2012fy}. It is numerically found that our predictions always lie within the margins of the Belle measurements, except for a $1.7~\sigma$ deviation for the lowest mass bin~\cite{Bischofberger:2011pw}. While being below the current Belle detection sensitivity that is of $\mathcal{O}(10^{-3})$, our predictions are expected to be detectable at the Belle II experiment~\cite{Kou:2018nap}, where $\sqrt{70}$ times more sensitive results will be available with a $50~\text{ab}^{-1}$ data sample.

Our paper is organized as follows. In Sec.~\ref{sec:ACP}, we firstly derive the $CP$-violating angular observable in $\tau\to K_S\pi\nu_\tau$ decays by means of the reciprocal basis. The angular distribution of $\tau^-\to \bar{K}^0\pi^-\nu_\tau$ decay is then presented in Sec.~\ref{sec:Angular-distribution}, and Sec.~\ref{sec:numerical} contains our numerical results and discussions. Finally, we conclude in Sec.~\ref{sec:conclusion}. For convenience, dispersive representations of the $K\pi$ vector and scalar form factors are given in the appendix.

\section{\boldmath $CP$-violating angular observable in $\tau\to K_S\pi\nu_\tau$ decay}
\label{sec:ACP}
\begin{figure}[ht]
	\centering
	\includegraphics[width=0.38\textwidth]{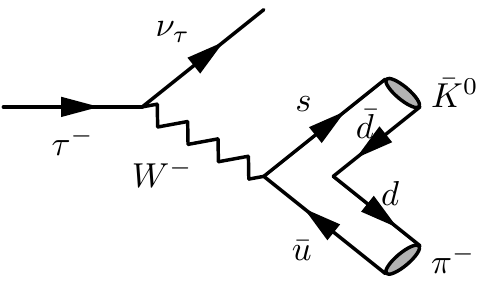}
	\hspace{0.45in}
	\includegraphics[width=0.38\textwidth]{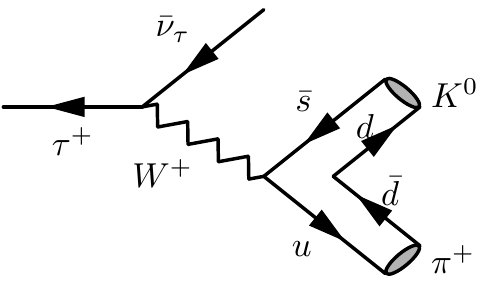}
	\caption{\label{fig:Feynman diagram} \small Tree-level Feynman diagrams for the decay $\tau^-\to \bar{K}^0\pi^-\nu_\tau$ (left) and its $CP$-conjugated mode $\tau^+\to K^0\pi^+\bar\nu_\tau$ (right) within the SM.}
\end{figure}

As the $\tau^-$~($\tau^+$) decay produces initially a $\bar{K}^0$~($K^0$) state due to the $\Delta S=\Delta Q$ rule, we have for the SM transition amplitudes the following relation: 
\begin{align}\label{eq:amp_sm}
\mathcal{A}(\tau^-\to\bar{K}^0\pi^-\nu_\tau)=\mathcal{A}(\tau^+\to K^0\pi^+\bar{\nu}_\tau)\,,
\end{align}
which is due to the fact that the CKM matrix element $V_{us}$ involved is real and the strong phase must be the same for these two $CP$-related processes~\cite{Bigi:2005ts}. The relevant Feynman diagrams at the tree level in weak interaction are shown in Fig.~\ref{fig:Feynman diagram}.

Due to the well-established $K^0-\bar{K}^0$ mixing~\cite{Christenson:1964fg,Tanabashi:2018oca}, the experimentally reconstructed kaons are not the flavour ($|K^0\rangle$ and $|\bar{K}^0\rangle$) but rather the mass ($|K_S\rangle$ and $|K_L\rangle$) eigenstates, which, in the absence of $CP$ violation in the system, are related to each other via~\cite{Branco:1999fs,Silva:2004gz}
\begin{align}\label{eq:mixing_no_cp}
|K_{S,L}\rangle=\frac{1}{\sqrt{2}}\left(|K^0\rangle \pm e^{i\zeta}|\bar{K}^0\rangle\right)\,,
\end{align}
where $\zeta$ is the spurious phase brought about by the $CP$ transformation, $\mathcal{CP}|K^0\rangle=e^{i\zeta}|\bar{K}^0\rangle$~\cite{Silva:2004gz}. In this case, the double differential decay widths of $\tau\to K_S\pi\nu_\tau$ decays satisfy
\begin{align}\label{eq:dwidth_no_cp}
\frac{d^2\Gamma(\tau^-\to K_{S,L}\pi^-\nu_\tau)}{ds\,d\cos\alpha}=\frac{1}{2}\,\frac{d^2\Gamma(\tau^-\to \bar{K}^0\pi^-\nu_\tau)}{ds\,d\cos\alpha}\,,\nonumber\\[0.2cm]
\frac{d^2\Gamma(\tau^+\to K_{S,L}\pi^+\bar\nu_\tau)}{ds\,d\cos\alpha}=\frac{1}{2}\,\frac{d^2\Gamma(\tau^+\to K^0\pi^+\bar\nu_\tau)}{ds\,d\cos\alpha}\,,
\end{align}
which, taken together with Eq.~\eqref{eq:amp_sm}, indicate that there exists \cp\ asymmetry neither in the integrated decay rate nor in the angular distribution of $\tau\to K_S\pi\nu_\tau$ decays within the SM.\footnote{Here we do not consider the contribution from second-order weak interaction, which is estimated to be of $\mathcal{O}(10^{-12})$, and can be therefore neglected safely~\cite{Delepine:2005tw}.}

Once the well-measured $CP$ violation in $K^0-\bar{K}^0$ mixing~\cite{Christenson:1964fg,Tanabashi:2018oca} is taken into account, however, the two mass eigenkets $|K_{S,L}\rangle$ will be now given by~\cite{Silva:2004gz}\footnote{The $CPT$ invariance is still assumed here. For the general case in which both $CP$ and $CPT$ are violated in the mixing, the readers are referred to Refs.~\cite{Branco:1999fs,Silva:2004gz}.}
\begin{align}\label{eq:mixing_cp}
|K_{S,L}\rangle=p\,|K^0\rangle \pm q\,|\bar{K}^0\rangle\,,
\end{align}
with the normalization $|p|^2+|q|^2=1$. The corresponding mass eigenbras  $\langle\tilde K_{S,L}|$ read~\cite{Silva:2004gz}
\begin{align}\label{eq:reciprocal}
\langle \tilde K_{S,L} |=\frac{1}{2}\left(p^{-1}\langle K^0| \pm q^{-1}\langle \bar{K}^0|\right)\,,
\end{align}
which form the so-called reciprocal basis that is featured by both the orthornormality and completeness conditions~\cite{Silva:2004gz}:
\begin{align}\label{eq:relation}
& \langle \tilde K_{S} |K_{S}\rangle=\langle \tilde K_{L} |K_{L}\rangle=1\,,\quad
\langle \tilde K_{S} |K_{L}\rangle=\langle \tilde K_{L} |K_{S}\rangle=0\,,\nonumber\\[0.2cm]
& \hspace{2.5cm} |K_{S}\rangle\langle \tilde K_{S} |+|K_{L}\rangle\langle \tilde K_{L} |=1\,.
\end{align} 
Notice that the mass eigenbras $\langle\tilde K_{S,L}|$ do not coincide with $\langle K_{S,L}|$, the Hermitian conjugates of the mass eigenkets $|K_{S,L}\rangle$. This is because the $2\times2$ effective Hamiltonian $\mbox{\boldmath $H$}$ responsible for the \kkb\ mixing is not a normal matrix, and hence cannot be diagonalized by a unitary transformation but rather by a general similarity transformation~\cite{Branco:1999fs,Silva:2000db,Silva:2004gz}. Consequently, the time-evolution operator for the \kkb\ system is determined by
\begin{equation}\label{eq:time-evolution}
\text{exp}(-i\mbox{\boldmath $H$}t)=e^{-i\mu_{S}t}|K_S\rangle\langle \tilde K_S|+e^{-i\mu_{L}t}|K_L\rangle\langle \tilde K_L|\,,
\end{equation}
where $\mu_{S,L}=M_{S,L}-i/2\,\Gamma_{S,L}$ are the two eigenvalues of the effective Hamiltonian $\mbox{\boldmath $H$}$, with the real and imaginary parts representing their masses and decay widths, respectively. 

Experimentally, the intermediate state $K_S$ in $\tau\to K_S\pi\nu_\tau$ decays is not directly observed, but rather reconstructed in terms of a $\pi^+\pi^-$ final state with $M_{\pi\pi}\approx M_K$~\cite{Anderson:1998ke,Bonvicini:2001xz,Bischofberger:2011pw,Epifanov:2007rf}. When $CP$ violation in the neutral kaon system is invoked, however, the final state $\pi^+\pi^-$ can be obtained not only from the short-lived $K_S$, but also from the long-lived $K_L$ state, when the kaon decay time is long enough. Thus, the processes $\tau^{\pm}\to [\pi^+\pi^-]\pi^\pm\bar{\nu}_\tau(\nu_\tau)$ proceed actually as follows: the initial states $\tau^\pm$ decay into the intermediate states $K_{S,L}\pi^\pm\bar{\nu}_\tau(\nu_\tau)$, which after a time $t$ decay into the final state $[\pi^+\pi^-]\pi^\pm\bar{\nu}_\tau(\nu_\tau)$. In this context, it is convenient to apply Eq.~\eqref{eq:time-evolution} to describe the time evolution of these processes. With the reference to $\pi^\pm\bar{\nu}_\tau(\nu_\tau)$ suppressed, the complete amplitudes for these two \cp-related processes can be written as~\cite{Silva:2000db,Chen:2019vbr}
\begin{align}
\mathcal{A}(\tau^-\to K_{S,L}\to\pi^+\pi^-)&=\langle \pi^+\pi^-|T|K_S\rangle e^{-i\mu_St}\langle \tilde K_S|T|\tau^-\rangle+\langle \pi^+\pi^-|T|K_L\rangle e^{-i\mu_Lt}\langle \tilde K_L|T|\tau^-\rangle\,\nonumber\\[0.2cm]
&=\frac{1}{2q}\Big[\langle \pi^+\pi^-|T|K_S\rangle e^{-i\mu_St}-\langle \pi^+\pi^-|T|K_L\rangle e^{-i\mu_Lt}\Big]\langle\bar K^0|T|\tau^-\rangle\,,\label{eq:amp1}\\[0.3cm]
\mathcal{A}(\tau^+\to K_{S,L}\to\pi^+\pi^-)&=\langle \pi^+\pi^-|T|K_S\rangle e^{-i\mu_St}\langle \tilde K_S|T|\tau^+\rangle+\langle \pi^+\pi^-|T|K_L\rangle e^{-i\mu_Lt}\langle \tilde K_L|T|\tau^+\rangle\,\nonumber\\[0.2cm]
&=\frac{1}{2p}\Big[\langle \pi^+\pi^-|T|K_S\rangle e^{-i\mu_St}+\langle \pi^+\pi^-|T|K_L\rangle e^{-i\mu_Lt}\Big]\langle K^0|T|\tau^+\rangle\,,\label{eq:amp2}
\end{align}
where Eq.~\eqref{eq:reciprocal} and the $\Delta S=\Delta Q$ rule have been used to obtain the second lines. It is obvious from Eqs.~\eqref{eq:amp1} and \eqref{eq:amp2} that the kaon decays are independent of the $\tau$ decays, which means that the complete double differential decay widths can be written as
\begin{align}
\frac{d^2\Gamma(\tau^-\to K_{S,L}\pi^-\nu_\tau\to[\pi^+\pi^-]\pi^-\nu_\tau)}{ds\,d\cos\alpha}=\frac{d^2\Gamma(\tau^-\to\bar K^0\pi^-\nu_\tau)}{ds\,d\cos\alpha}\,\Gamma(\bar K^0(t)\to\pi^+\pi^-)\,,\label{eq:evolution1}\\[0.2cm]
\frac{d^2\Gamma(\tau^+\to K_{S,L}\pi^+\bar\nu_\tau\to[\pi^+\pi^-]\pi^+\bar\nu_\tau)}{ds\,d\cos\alpha}=\frac{d^2\Gamma(\tau^+\to K^0\pi^+\bar\nu_\tau)}{ds\,d\cos\alpha}\,\Gamma(K^0(t)\to\pi^+\pi^-)\,,\label{eq:evolution2}
\end{align}
with the time-dependent $K\to\pi^+\pi^-$ decay widths given by
\begin{align}
\Gamma(\bar K^0(t)\to\pi^+\pi^-)=\frac{|\langle \pi^+\pi^-| T| K_S\rangle|^2}{4| q|^2}\Big[e^{-\Gamma_St}\!+\!|\eta_{+-}|^2\,e^{-\Gamma_Lt}\!-\!2|\eta_{+-}|\,e^{-\Gamma t}\cos(\phi_{+-}\!-\!\Delta m t) \Big]\,,\label{eq:K0bar}\\
\Gamma(K^0(t)\to\pi^+\pi^-)=\frac{|\langle \pi^+\pi^-| T| K_S\rangle|^2}{4| p|^2}\Big[e^{-\Gamma_St}\!+\!|\eta_{+-}|^2\,e^{-\Gamma_Lt}\!+\!2|\eta_{+-}|\,e^{-\Gamma t}\cos(\phi_{+-}\!-\!\Delta m t) \Big]\,,\label{eq:K0}
\end{align}
where $\Delta m=M_L-M_S$ and $\Gamma=\frac{\Gamma_L+\Gamma_S}{2}$ denote respectively the mass difference and the average width of the \kkb\ system, while $\eta_{+-}$ is defined as the $CP$-violating amplitude ratio for the $\pi^+\pi^-$ final state,
\begin{equation}
\eta_{+-}=\frac{\langle \pi^+\pi^-| T| K_L\rangle}{\langle \pi^+\pi^-| T| K_S\rangle}\,,
\end{equation}
with its modulus $|\eta_{+-}|=(2.232\pm 0.011)\times 10^{-3}$ and its phase $\phi_{+-}=(43.51\pm 0.05)^\circ$~\cite{Tanabashi:2018oca}. 

Keeping in mind that~\cite{Branco:1999fs,Tanabashi:2018oca}
\begin{align}
\frac{|p|^2-|q|^2}{|p|^2+|q|^2}=\frac{2\Re e(\epsilon_K)}{1+|\epsilon_K|^2}\approx 2\Re e(\epsilon_K)=(3.32\pm0.06)\times 10^{-3}\,, 
\end{align}
where $\epsilon_K$ is the \cp-violating parameter in neutral kaon decays, one can see from Eqs.~\eqref{eq:K0} and \eqref{eq:K0bar} that, for the sum of the two time-dependent decay widths, both the interference (the last) and the pure $K_L$ term (the second) are suppressed compared to the pure $K_S$ contribution (the first term in the square bracket); for their difference, however, the interference between the amplitudes of $K_S$ and $K_L$ is found to be as important as the pure $K_S$ amplitude~\cite{Grossman:2011zk}. As a consequence, the \cp-violating angular observable defined by Eq.~\eqref{eq:ACP_i} will depend on the times over which the differential decay rates are integrated. In addition, once the time evolution of the kaons are considered, one has to take into account not only the efficiency as a function of the kaon decay time, but also the kaon energy in the laboratory frame to account for the time dilation~\cite{Grossman:2011zk}. With all these experiment-dependent effects parametrized by a function $F(t)$~\cite{Grossman:2011zk}, and for a decay-time interval $[t_1,t_2]$, we can then define
\begin{align}\label{eq:ACPi}
A_i^{CP}(t_1,t_2)
&=\frac{\int_{s_{1,i}}^{s_{2,i}}\int_{-1}^{1} \cos \alpha \left[\frac{d \Gamma^{\tau^-}}{d\omega}\int_{t_1}^{t_2}F(t)\bar\Gamma_{\pi^+\pi^-}(t)\,dt -\frac{d \Gamma^{\tau^+}}{d\omega}\int_{t_1}^{t_2}F(t)\Gamma_{\pi^+\pi^-}(t)\,dt\right] d\omega}{\frac{1}{2}\int_{s_{1,i}}^{s_{2,i}}\int_{-1}^{1}\left[\frac{d \Gamma^{\tau^-}}{d\omega}\int_{t_1}^{t_2}F(t)\bar\Gamma_{\pi^+\pi^-}(t)\,dt +\frac{d \Gamma^{\tau^+}}{d\omega}\int_{t_1}^{t_2}F(t)\Gamma_{\pi^+\pi^-}(t)\,dt\right] d\omega}\nonumber\\[0.2cm]
&=\frac{\left(\langle\cos\alpha\rangle_i^{\tau^-}+\langle\cos\alpha\rangle_i^{\tau^+}\right)A^{CP}_K(t_1,t_2)+\left(\langle\cos\alpha\rangle_i^{\tau^-}-\langle\cos\alpha\rangle_i^{\tau^+}\right)}{1+A^{CP}_K(t_1,t_2)\cdot A_{\tau,i}^{CP}}\,,
\end{align}
with
\begin{align}
\langle\cos\alpha\rangle_i^{\tau^-}+\langle\cos\alpha\rangle_i^{\tau^+}&=\frac{\int_{s_{1,i}}^{s_{2,i}}\int_{-1}^{1} \cos \alpha \left[\frac{d \Gamma^{\tau^-}}{d\omega}+\frac{d \Gamma^{\tau^+}}{d\omega}\right]d\omega}{\frac{1}{2}\int_{s_{1,i}}^{s_{2,i}}\int_{-1}^{1} \left[\frac{d \Gamma^{\tau^-}}{d\omega}+\frac{d \Gamma^{\tau^+}}{d\omega}\right]d\omega}\,,\\[0.2cm]
\langle\cos\alpha\rangle_i^{\tau^-}-\langle\cos\alpha\rangle_i^{\tau^+}&=\frac{\int_{s_{1,i}}^{s_{2,i}}\int_{-1}^{1} \cos \alpha \left[\frac{d \Gamma^{\tau^-}}{d\omega}-\frac{d \Gamma^{\tau^+}}{d\omega}\right]d\omega}{\frac{1}{2}\int_{s_{1,i}}^{s_{2,i}}\int_{-1}^{1} \left[\frac{d \Gamma^{\tau^-}}{d\omega}+\frac{d \Gamma^{\tau^+}}{d\omega}\right]d\omega}\,,\\[0.2cm]
A^{CP}_K(t_1,t_2)&=\frac{\int_{t_1}^{t_2}F(t)\left[\bar\Gamma_{\pi^+\pi^-}(t)-\Gamma_{\pi^+\pi^-}(t)\right]dt}{\int_{t_1}^{t_2}F(t)\left[\bar\Gamma_{\pi^+\pi^-}(t)+\Gamma_{\pi^+\pi^-}(t)\right]dt}\,,\label{eq:AcpK}\\[0.2cm]
A_{\tau,i}^{CP}&=\frac{\int_{s_{1,i}}^{s_{2,i}}\int_{-1}^{1} \left[\frac{d \Gamma^{\tau^-}}{d\omega}-\frac{d \Gamma^{\tau^+}}{d\omega}\right]d\omega}{\int_{s_{1,i}}^{s_{2,i}}\int_{-1}^{1} \left[\frac{d \Gamma^{\tau^-}}{d\omega}+\frac{d \Gamma^{\tau^+}}{d\omega}\right]d\omega}\,,\label{eq:Acptau}
\end{align}
where $d\omega=ds\,d\cos\alpha$, $\frac{d\Gamma^{\tau^\pm}}{d\omega}=\frac{d^2\Gamma(\tau^\pm\to K^0(\bar K^0)\pi^\pm\bar\nu_\tau(\nu_\tau))}{ds\,d\cos\alpha}$, $\Gamma(\bar\Gamma)_{\pi^+\pi^-}(t)=\Gamma(K^0(\bar K^0)(t)\to\pi^+\pi^-)$, and $\langle\cos\alpha\rangle_i^{\tau^{\pm}}$ denote the differential $\tau^{\pm}$ decay widths weighted by $\cos\alpha$ and evaluated in the $i$-th bin. Within the SM, one has $A_{\tau,i}^{CP}=0$ and $\langle\cos\alpha\rangle_i^{\tau^-}=\langle\cos\alpha\rangle_i^{\tau^+}$ due to $\frac{d \Gamma^{\tau^+}}{d\omega}=\frac{d \Gamma^{\tau^-}}{d\omega}$, and thus the $CP$-violating angular observable defined by Eq.~\eqref{eq:ACPi} reduces to
\begin{align}\label{eq:reduceACPi}
A_i^{CP}(t_1,t_2) = 2\,\langle\cos\alpha\rangle_i^{\tau^-}A^{CP}_K(t_1,t_2)\,,
\end{align}
which is the key result obtained in this work, and indicates that, once the well-measured $CP$ violation in the neutral kaon system is invoked, a non-zero $CP$ asymmetry would appear in the angular observable of the decays considered, even within the SM.

As indicated by Eq.~\eqref{eq:reduceACPi}, in order to get a prediction of the \cp\ asymmetry $A_i^{CP}(t_1,t_2)$, one should firstly determine both $\langle\cos\alpha\rangle_i^{\tau^-}$ and $A^{CP}_K(t_1,t_2)$. The mean value of the angular observable $\langle\cos\alpha\rangle_i^{\tau^-}$ is related to the so-called forward-backward asymmetry $A_\text{FB}$~\cite{Beldjoudi:1994hi}, the computation of which will be detailed in the next section. As the observable $A^{CP}_K(t_1,t_2)$ is sensitive to the experimental cuts, its prediction can be made only when the kaon decay time interval over which it is integrated as well as the experiment-dependent function $F(t)$ have been determined. Here we shall quote the particularly simple prediction made in Ref.~\cite{Grossman:2011zk}, 
\begin{equation}\label{eq:ACPK}
A^{CP}_K(t_1\ll\Gamma_S^{-1},\Gamma_S^{-1}\ll t_2 \ll\Gamma_L^{-1})\approx -2\Re e(\epsilon_K)=-3.32\times10^{-3}\,,
\end{equation}
in which the approximations with $|\eta_{+-}|\approx \frac{2\Re e(\epsilon_K)}{\sqrt{2}}$, $\phi_{+-}\approx 45^\circ$, $\Gamma\approx\frac{\Gamma_S}{2}$, and $\Delta m\approx\frac{\Gamma_S}{2}$~\cite{Branco:1999fs}, as well as a double step function~\cite{Grossman:2011zk}
\begin{align}\label{eq:doublestep}
F(t)=\left\{
\begin{aligned}
1, & & \mbox{$t_1<t<t_2$}\\[0.2cm]
0, & & \mbox{otherwise}
\end{aligned}
\right.\,,
\end{align}
have been used in Eqs.~\eqref{eq:K0bar}, \eqref{eq:K0} and \eqref{eq:AcpK}. 

It should be noted that, when using the simple double-step form of $F(t)$ given by Eq.~\eqref{eq:doublestep}, one has assumed that the $K_S$ state can be fully reconstructed within the time interval $[t_1,t_2]$, with $t_1\ll\Gamma_S^{-1}$ and $\Gamma_S^{-1}\ll t_2 \ll \Gamma_L^{-1}$~\cite{Grossman:2011zk}. This might be, however, not always the case in experiment. For example, a different parametrization of $F(t)$ has been used by the BaBar collaboration~\cite{BABAR:2011aa}, resulting in a multiplicative correction factor of $1.08\pm0.01$ for the observable $A^{CP}_K(t_1,t_2)$ given by Eq.~\eqref{eq:ACPK}. Due to different experimental conditions, different forms of $F(t)$ could also be adopted by the Belle and Belle II collaborations. Thus, our theoretical predictions given by Eqs.~\eqref{eq:ACPK} and \eqref{eq:extra}, as well as in Table~\ref{tab:results} should be refined once the explicit forms of the function $F(t)$ are ultimately determined by the experimental collaboration.

\section{\boldmath Angular distribution in $\tau^-\to \bar K^0\pi^-\nu_\tau$ decay}
\label{sec:Angular-distribution}
Within the SM, the effective weak Hamiltonian responsible for the strangeness-changing hadronic $\tau$ decays is given by
\begin{align}\label{eq:Heff}
\mathcal{H}_{\text{eff}}=\frac{G_F}{\sqrt{2}}\,V_{us}\left[\bar{\tau}\gamma_\mu(1-\gamma_5)\nu_\tau\right]\left[\bar{u}\gamma^\mu(1-\gamma_5)s\right]+\mathrm{h.c.}\,,
\end{align}   
where $G_{F}$ is the Fermi coupling constant, and $V_{us}$ is the CKM matrix element involved in the transitions. The decay amplitude for $\tau^-\to\bar{K}^0\pi^-\nu_\tau$ decay can then be written as
\begin{align}\label{eq:amp}
\mathcal{A}(\tau^-\to\bar{K}^0\pi^-\nu_\tau)=\frac{G_F V_{us}\sqrt{S_{\text{EW}}}}{\sqrt{2}}\,L_\mu\,H^\mu\,,
\end{align}
where $S_{\text{EW}}=1.0201(3)$ encodes the short-distance electroweak radiative correction to the hadronic $\tau$ decays~\cite{Erler:2002mv}. In Eq.~\eqref{eq:amp}, $L_\mu$ denotes the leptonic current given by
\begin{align}
L_\mu=\bar{u}(p_{\nu_\tau})\gamma_\mu(1-\gamma_5)u(p_\tau)\,,
\end{align}
while $H^\mu$ denotes the hadronic matrix element and can be parametrized as\footnote{Due to parity conservation in strong interaction, the hadronic matrix element for a transition from the vacuum to two pseudo-scalar mesons involves the vector current only.}
\begin{align}
H^\mu=\left\langle \bar{K}^0(p_K)\pi^-(p_\pi)\left|\bar{s}\gamma^\mu u\right|0\right\rangle=\left[(p_K-p_\pi)^\mu-\frac{\Delta_{K\pi}}{s}q^\mu\right]F_+(s)+\frac{\Delta_{K\pi}}{s}q^\mu F_0(s)\,,
\end{align}
where $s=(p_K+p_\pi)^2$, $q^\mu=(p_K+p_\pi)^\mu$, $\Delta_{K\pi}=M_K^2-M_{\pi}^2$, and $F_+(s)$ and $F_0(s)$ are the $K\pi$ vector and scalar form factors associated with the $J^P=1^-$ and $J^P=0^+$ components of the weak charged current, respectively. As mentioned already in Sec.~\ref{sec:intro}, we shall use the dispersive representations rather than the Breit-Wigner parameterizations of these form factors in this work. For convenience, their explicit expressions are presented in the appendix. 

\begin{figure}[t]
	\centering
	\includegraphics[width=0.475\textwidth]{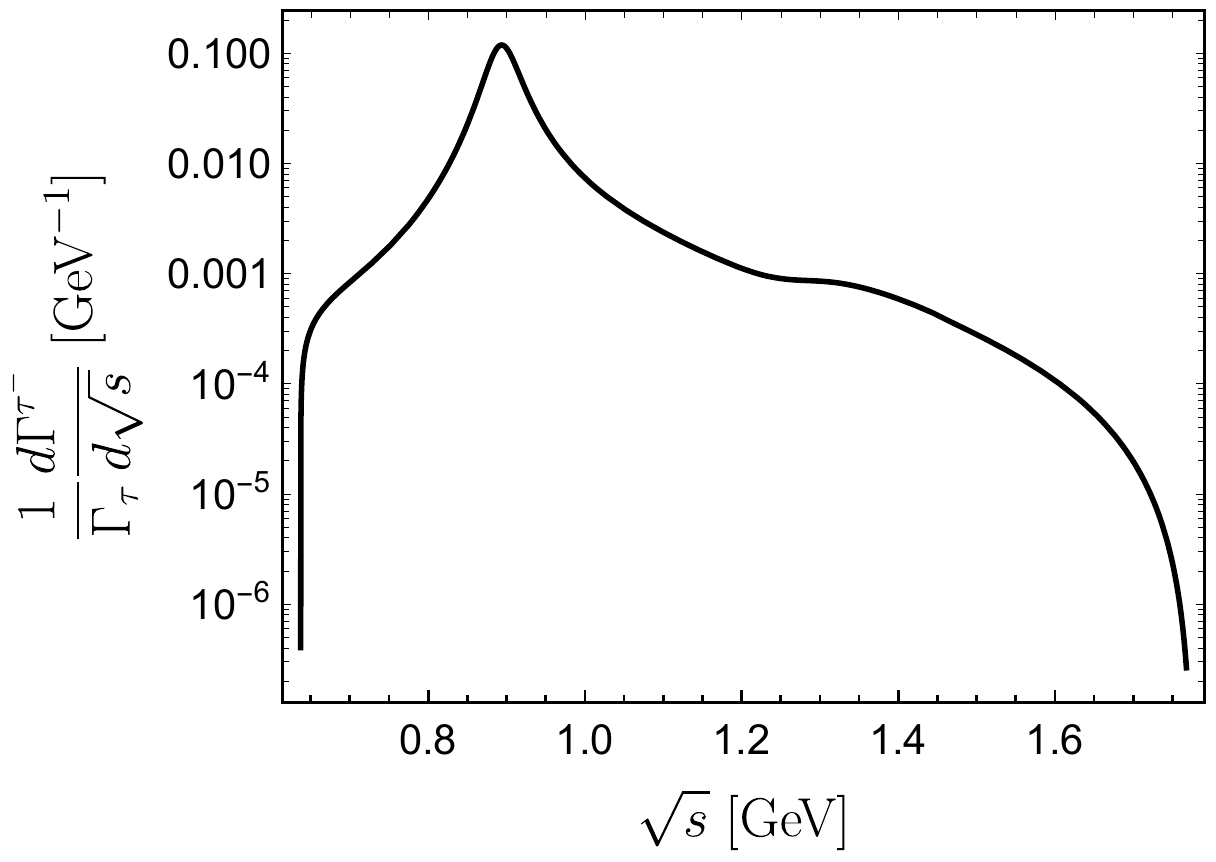}
	\hspace{0.45in}
	\includegraphics[width=0.44\textwidth]{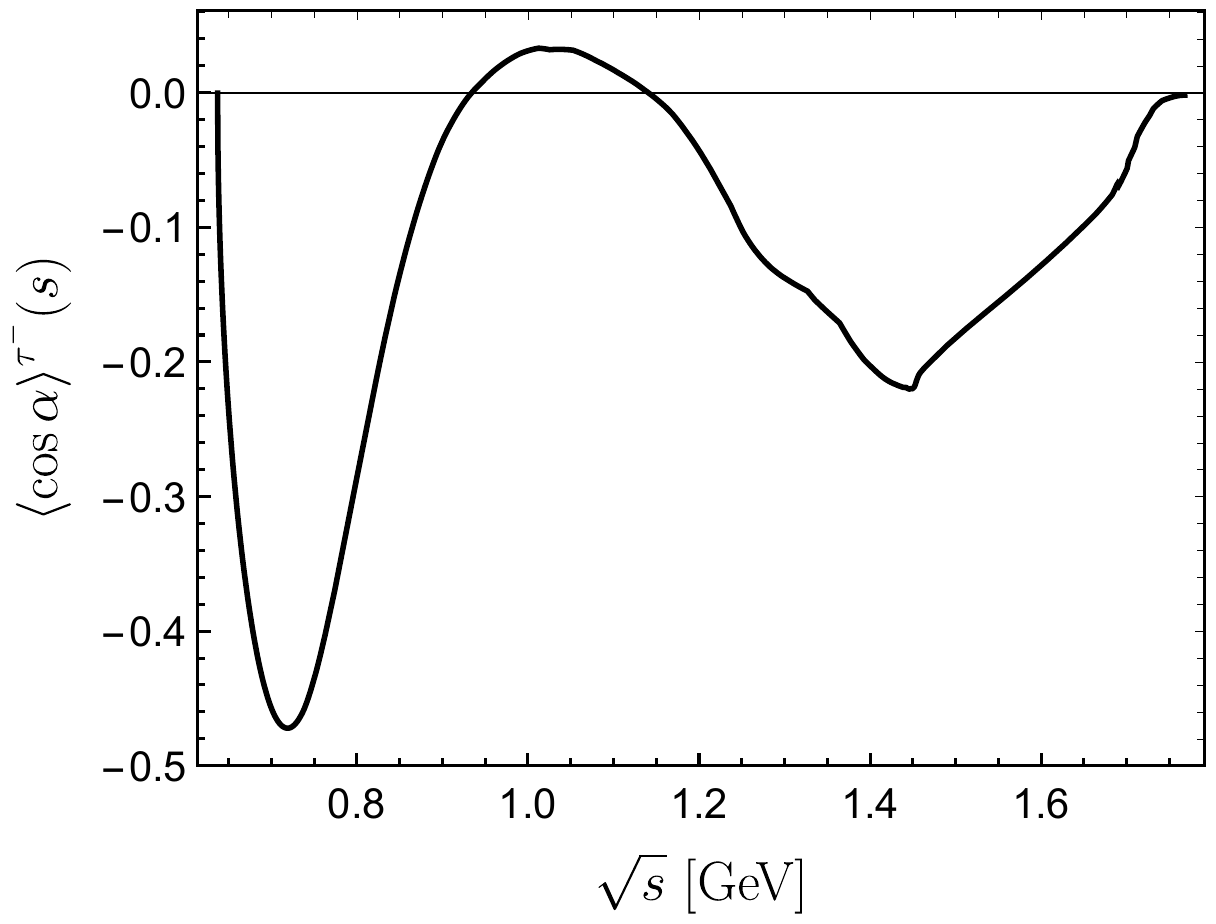}
	\caption{\label{fig:observables} \small The $K\pi$ invariant mass distribution of the differential decay width $\frac{d\Gamma^{\tau^-}}{d\sqrt{s}}$ normalized by the $\tau$ total decay width (left) as well as the angular observable $\langle\cos\alpha\rangle^{\tau^-}(s)$ (right) for the decay  $\tau^-\to\bar{K}^0\pi^-\nu_\tau$ within the SM.}
\end{figure}

Working in the $K\pi$ rest frame, and after integrating over the unobserved neutrino direction, one can then write the double differential decay width of $\tau^-\to\bar{K}^0\pi^-\nu_\tau$ decay as~\cite{Kuhn:1992nz,Kuhn:1996dv}
\begin{align}\label{differetial_width}
\frac{d^2\Gamma^{\tau^-}}{ds\,d\cos\alpha}=&\frac{G_F^2| F_+(0)V_{us}|^2m_\tau^3 S_{\rm EW}}{512\pi^3s^3}\left(1-\frac{s}{m_\tau^2}\right)^2\lambda^{1/2}(s,M_K^2,M_{\pi}^2)\nonumber\\
&\times\Bigg\lbrace\Big|\tilde{F}_+(s)\Big|^2\left(\frac{s}{m_\tau^2}+(1-\frac{s}{m_\tau^2})\cos^2\alpha\right)\lambda(s,M_K^2,M_{\pi}^2)+\Delta_{K\pi}^2\Big|\tilde{F}_0(s)\Big|^2\nonumber\\
&\hspace{0.8cm}-2\Delta_{K\pi}\Re e\big[\tilde{F}_+(s)\tilde{F}_0^*(s)\big]\lambda^{1/2}(s,M_K^2,M_{\pi}^2)\cos\alpha\Bigg\rbrace\,,
\end{align}
where $\tilde{F}_{+,0}(s)=F_{+,0}(s)/F_+(0)$, and $\lambda(s,M_K^2,M_{\pi}^2)=\left[s-(M_K+M_{\pi})^2\right]\left[s-(M_K-M_{\pi})^2\right]$. Integrating Eq.~\eqref{differetial_width} over $\cos\alpha$, one then arrives at the differential decay width as a function of the $K\pi$ invariant mass squared,
\begin{align}\label{eq:spectrum}
\frac{d\Gamma^{\tau^-}}{ds}
=&\frac{G_{F}^{2}|F_+(0)V_{us}|^{2} m_{\tau}^{3} S_{\text{EW}}}{512 \pi^{3} s^3}\left(1-\frac{s}{m_{\tau}^{2}}\right)^{2} \lambda^{1/2}(s, M_K^{2}, M_\pi^{2})\nonumber\\
&\times\Bigg\lbrace \frac{2}{3}\lambda(s, M_K^{2}, M_\pi^{2})\left(1+\frac{2s}{m_\tau^2}\right)\Big| \tilde{F}_+(s)\Big|^2+2\Delta_{K\pi}^2\Big| \tilde{F}_0(s)\Big|^2\Bigg\rbrace\,,
\end{align}
which involves only the moduli of the two $K\pi$ form factors, and is dominated by the vector one. This implies that there exists no \cp\ violation in the decay rate within the SM~\cite{Bigi:2005ts,Calderon:2007rg,Grossman:2011zk}. The $K\pi$ invariant mass distribution of the differential decay width $\frac{d\Gamma^{\tau^-}}{d\sqrt{s}}$ normalized by the $\tau$ total decay width is shown in the left plot of Fig.~\ref{fig:observables}, from which a clear peak structure at the vicinity of $\sqrt{s}\sim0.9~\text{GeV}$ and a mild bump at the vicinity of $\sqrt{s}\sim1.4~\text{GeV}$ can be seen, indicating the existence of $K^*(892)$ and $K^*(1410)$ resonances, respectively.

To obtain further information about the $K\pi$ vector and scalar form factors and, especially, about their relative phase that is of particular interest in relation to the study of \cp\ violation, one must resort to other observables that involve the interference between these two form factors. For this purpose, one introduces the angular observable~\cite{Kou:2018nap}
\begin{align}\label{eq:averagecosalpha}
\langle\cos\alpha\rangle^{\tau^{-}}(s)&= \frac{\int_{-1}^{1} \cos \alpha \left(\frac{d^2\Gamma^{\tau^-}}{ds\,d\cos\alpha}\right) d\cos \alpha}{\int_{-1}^{1} \left(\frac{d^2 \Gamma^{\tau^-}}{ds\,d\cos\alpha}\right) d\cos\alpha}\nonumber\\[0.1cm]
&=\frac{-2\Delta_{K\pi}\Re e[\tilde{F}_+(s)\tilde{F}_0^*(s)]\lambda^{1/2}\left(s, M_{K}^{2}, M_{\pi}^{2}\right)}{\Big| \tilde{F}_+(s)\Big|^2\left(1+\frac{2s}{m_\tau^2}\right)\lambda\left(s, M_{K}^{2}, M_{\pi}^{2}\right)+3\Delta_{K\pi}^2\Big| \tilde{F}_0(s)\Big|^2}\,,
\end{align}
which is defined as the differential decay width weighted by $\cos\alpha$. It is interesting to note that the observable $\langle\cos\alpha\rangle^{\tau^{-}}(s)$ is connected with the so-called forward-backward asymmetry $A_\text{FB}(s)$ via the relation $\langle\cos \alpha\rangle^{\tau^{-}}(s)=\frac{2}{3}A_\text{FB}^{\tau^-}(s)$~\cite{Kou:2018nap}, with the latter defined by~\cite{Beldjoudi:1994hi,Kimura:2014wsa,Gao:2012su}
\begin{align}\label{eq:AFB}
A_\text{FB}^{\tau^-}(s)&=\frac{\int_0^1\frac{d^2 \Gamma^{\tau^-}}{ds\,d\cos\alpha}d\cos\alpha-\int_{-1}^0\frac{d^2 \Gamma^{\tau^-}}{ds\,d\cos\alpha}d\cos\alpha}{\int_0^1\frac{d^2 \Gamma^{\tau^-}}{ds\,d\cos\alpha}d\cos\alpha+\int_{-1}^0\frac{d^2  \Gamma^{\tau^-}}{ds\,d\cos\alpha}d\cos\alpha}\,.
\end{align}
Being proportional to the factor $\Delta_{K\pi}=M_K^2-M_{\pi}^2$ that would vanish in the limit of the exact $\text{SU(3)}$ flavour symmetry, the angular observable $\langle\cos \alpha\rangle^{\tau^{-}}(s)$ (or equivalently the forward-backward asymmetry $A_\text{FB}^{\tau^-}(s)$) also allows us to measure the $\text{SU(3)}$ breaking effect in the decays considered~\cite{Beldjoudi:1994hi}. The $K\pi$ invariant mass distribution of $\langle\cos\alpha\rangle^{\tau^{-}}(s)$ is shown in the right plot of Fig.~\ref{fig:observables}, from which two negative extrema are observed at the vicinities of $\sqrt{s}\sim 0.72~\text{GeV}$ and $\sqrt{s}\sim 1.45~\text{GeV}$. This indicates that a non-zero  $\langle\cos\alpha\rangle^{\tau^{-}}(s)$ (or equivalently $A_\text{FB}(s)$) can be measured in this channel around $\sqrt{s}\sim 0.72~\text{GeV}$.

As the \cp-violating angular observable $A_i^{CP}(t_1,t_2)$ defined by Eq.~\eqref{eq:reduceACPi} is usually measured in bins of the $K\pi$ invariant mass~\cite{Anderson:1998ke,Bonvicini:2001xz,Bischofberger:2011pw}, one can also make the observable $\langle\cos\alpha\rangle^{\tau^-}(s)$ to be bin-dependent, 
\begin{align}\label{eq:cosbin}
\langle\cos \alpha\rangle^{\tau^{-}}_i =& \frac{\int_{s_{1,i}}^{s_{2,i}}\int_{-1}^{1} \cos \alpha \left(\frac{d^2\Gamma^{\tau^-}}{ds\,d\cos\alpha}\right)ds\,d\cos \alpha}{\int_{s_{1,i}}^{s_{2,i}}\int_{-1}^{1} \left(\frac{d^2 \Gamma^{\tau^-}}{ds\,d\cos\alpha}\right) ds\,d\cos\alpha}\,.
\end{align}
The right plot of Fig.~\ref{fig:observables} suggests then that, in order to obtain a value of $\langle\cos \alpha\rangle^{\tau^{-}}_i$ as large as possible, the $K\pi$ invariant mass bins can be optimally set at the vicinities of the two negative extrema of $\langle\cos \alpha\rangle^{\tau^{-}}(s)$. To see this clearly, we shall make a detailed numerical estimate in the next section.

\section{Numerical results and discussions}
\label{sec:numerical}
\begin{table}[t]
	\tabcolsep 0.10in
	\renewcommand\arraystretch{1.8}
	\begin{center}
		\caption{\label{tab:input} \small Summary of the input parameters used throughout this work.}
		\vspace{0.18cm}
		\begin{tabular}{|c|c|c|c|c|}
			\hline\hline
			\multicolumn{5}{|l|}{QCD and electroweak parameters}\\
			\hline
			$G_F[10^{-5}~\text{GeV}^{-2}]$~\cite{Tanabashi:2018oca} & $S_{\rm EW}$~\cite{Erler:2002mv} & $\left|V_{us}F_+(0)\right|$~\cite{Moulson:2017ive} & $F_{\pi}~[\text{MeV}]$~\cite{Tanabashi:2018oca}  & $F_K~[\text{MeV}]$~\cite{Tanabashi:2018oca}\\
			\hline
			$1.1663787(6)$ & $1.0201(3)$ & $0.21654(41)$ & $92.3(1)$  & $1.198F_\pi$\\
			\hline
			\multicolumn{5}{|l|}{Particle masses and the $\tau$ lifetime~\cite{Tanabashi:2018oca}}\\
			\hline
			$m_{\tau}~[\text{MeV}]$ & $M_{K^0}~[\text{MeV}]$ & $M_{\pi^-}~[\text{MeV}]$ & \multicolumn{2}{c|}{$\tau_\tau~[10^{-15}~\text{s}]$}\\
			\hline
			$1776.86$ & $497.61$ & $139.57$ & \multicolumn{2}{c|}{$290.3$}\\
			\hline
			\multicolumn{5}{|l|}{Parameters in the $K\pi$ vector form factor with $s_{cut}=4~\text{GeV}^2$~\cite{Boito:2008fq}}\\
			\hline
			$m_{K^*}~[\text{MeV}]$ & $\gamma_{K^*}~[\text{MeV}]$ & $m_{K^{*\prime}}~[\text{MeV}]$ & $\gamma_{K^{*\prime}}~[\text{MeV}]$& $\gamma$\\
			\hline
			$943.41\pm0.59$& $66.72\pm0.87$ & $1374\pm45$ & $240\pm131$ & $-0.039\pm0.020$\\
			\hline
			$M_{K^*}~[\text{MeV}]$ &\multicolumn{2}{c|}{$\lambda_+^{\prime}$} & \multicolumn{2}{c|}{$\lambda_+^{\prime\prime}$ }\\
			\hline
			$892.01\pm0.92$&\multicolumn{2}{c|}{$(24.66\pm0.77)\times10^{-3}$} & \multicolumn{2}{c|}{$(11.99\pm0.20)\times10^{-4}$}\\
			\hline
			\multicolumn{5}{|l|}{$CP$-violating parameters in the neutral kaon system~\cite{Tanabashi:2018oca} }\\
			\hline
			$\left|\eta_{+-}\right|\times10^{3}$ & \multicolumn{2}{|c|}{$\phi_{+-}$} &\multicolumn{2}{|c|}{$\Re e(\epsilon_K)\times10^{3}$}\\
			\hline
			$2.232\pm0.011$ & \multicolumn{2}{|c|}{$(43.51\pm0.05)^{\circ}$} & \multicolumn{2}{|c|}{$1.66\pm0.02$}\\
			\hline \hline
		\end{tabular}
	\end{center}
\end{table}

Before presenting our numerical results, we firstly collect in Table~\ref{tab:input} all the input parameters used throughout this work. For any further details, the readers are referred to the original references. With the time-dependent \cp-violating observable $A^{CP}_K(t_1,t_2)$ fixed by Eq.~\eqref{eq:ACPK}, the computation of the \cp-violating angular observable $A_i^{CP}(t_1,t_2)$ is then attributed to that of the observable $\langle\cos\alpha\rangle^{\tau^{-}}_i$, which becomes now straightforward.

\begin{table}[ht]
	\tabcolsep 0.58in
	\renewcommand\arraystretch{1.8}
	\begin{center}
		\caption{\label{tab:results} \small Our predictions within the SM as well as the Belle measurements for the \cp-violating angular observable $A_i^{CP}(t_1,t_2)$ in four different mass bins.}
		\vspace{0.18cm}
		\begin{tabular}{ccc}
			\hline\hline
			$\sqrt{s}$~[GeV] & SM~$[10^{-3}]$ & Belle~$[10^{-3}]$~\cite{Bischofberger:2011pw}\\
			\hline
			$0.625-0.890$ & $0.80\pm0.02$ & $7.9\pm3.0\pm2.8$\\
			$0.890-1.110$ & $0.09\pm0.01$ & $1.8\pm2.1\pm1.4$\\
			$1.110-1.420$ & $0.50\pm0.09$ & $-4.6\pm7.2\pm1.7$\\
			$1.420-1.775$ & $1.19\pm0.24$ & $-2.3\pm19.1\pm5.5$\\
			\hline\hline
		\end{tabular}
	\end{center}
\end{table}

In order to make a direct numerical comparison with the Belle measurements, we choose the same intervals for the four bins of the $K\pi$ invariant mass as in Ref.~\cite{Bischofberger:2011pw}. Notice that the mass threshold used by the Belle collaboration for the lowest mass bin,  $s_{K\pi}=0.625~\text{GeV}$~\cite{Bischofberger:2011pw}, lies slightly  below the theoretical one, $s_{K\pi}=M_{K}+M_{\pi}=0.637~\text{GeV}$. As such a numerical difference has only a marginal impact on our prediction, we shall use the latter as input in this work. Our final predictions for the \cp-violating angular observable $A_i^{CP}(t_1,t_2)$ are shown in Table~\ref{tab:results}, where, for a comparison, the Belle measurements after background subtraction in each mass bin have also been given. One can see that our predictions always lie within the margins of the Belle results~\cite{Bischofberger:2011pw}, except for a $1.7~\sigma$ deviation for the lowest mass bin. It should be pointed out that our predictions, while being below the current Belle detection sensitivity that is of $\mathcal{O}(10^{-3})$, are expected to be detectable at the Belle II experiment~\cite{Kou:2018nap}, where $\sqrt{70}$ times more sensitive results will be obtained following the increase of the integrated luminosity of a $50~\text{ab}^{-1}$ data sample.

As mentioned already in last section, in order to get a value of $A_i^{CP}(t_1,t_2)$ as large as possible for a given time interval, we present two more predictions with the $K\pi$ invariant mass intervals selected at the vicinities of the two negative extrema of $\langle\cos\alpha\rangle^{\tau^{-}}$:
\begin{align}\label{eq:extra}
A_i^{CP}(t_1,t_2)=\left\{
\begin{aligned}
(3.06\pm0.06)\times10^{-3}, & & 0.70~\text{GeV}<\sqrt{s}< 0.75~\text{GeV}\\
(1.38\pm0.18)\times10^{-3}, & & 1.40~\text{GeV}<\sqrt{s}< 1.50~\text{GeV}
\end{aligned}
\right.\,.
\end{align}
It is interesting to note that the value of this observable in the mass interval $[0.70,0.75]~\text{GeV}$ is as large as the SM prediction for the decay-rate asymmetry~\cite{Grossman:2011zk}. Thus, we suggest the experimental $\tau$ physics groups at Belle II to measure the \cp-violating angular observable in this mass interval.   

\section{Conclusion}
\label{sec:conclusion}
In this work, inspired by the study of decay-rate asymmetry in $\tau\to K_S\pi\nu_\tau$ decays induced by the known \cp\ violation in \kkb\ mixing~\cite{Bigi:2005ts,Grossman:2011zk}, we have performed an investigation of the same effect on the \cp\ asymmetry in the angular distribution of the same channels within the SM. Our main conclusions are summarized as follows:
\begin{enumerate}[(i)]
\item Once the well-measured $CP$ violation in the neutral kaon system is invoked, a non-zero $CP$ asymmetry would appear in the angular observable of the decays considered, even within the SM. By utilizing the reciprocal basis, which has been used to reproduce conveniently the decay-rate asymmetry~\cite{Chen:2019vbr}, this observable is derived to be two times the product of the time-dependent $CP$ asymmetry in $K\to \pi^+\pi^-$ and the mean value of the angular distribution in $\tau^\pm\to K^0(\bar{K}^0)\pi^\pm\bar{\nu}_\tau(\nu_\tau)$ decays (see Eq.~\eqref{eq:reduceACPi}).

\item As the relative phase between the $K\pi$ vector and scalar form factors is required for the study of \cp\ violation, but the form-factor phases fitted via a superposition of Breit-Wigner functions do not vanish at threshold and violate Watson's theorem before the higher resonances start to play an effect~\cite{Cirigliano:2017tqn}, we did not adopt the Breit-Wigner parameterizations of these form factors. Instead, the thrice-subtracted (for the vector form factor)~\cite{Boito:2008fq,Boito:2010me} and the coupled-channel (for the scalar form factor)~\cite{Jamin:2000wn,Jamin:2001zq,Jamin:2006tj} dispersive representations have been employed, which warrant the properties of unitarity and analyticity, and contain a full knowledge of QCD in both the perturbative and non-perturbative regimes.

\item Our predictions for the \cp-violating angular observable $A_i^{CP}(t_1,t_2)$ always lie within the margins of the Belle measurements~\cite{Bischofberger:2011pw}, except for a $1.7~\sigma$ deviation for the lowest mass bin. While being below the current Belle detection sensitivity that is of $\mathcal{O}(10^{-3})$, our predictions are expected to be detectable at the Belle II experiment~\cite{Kou:2018nap}, where $\sqrt{70}$ times more sensitive results will be obtained following the increase of the integrated luminosity of a $50~\text{ab}^{-1}$ data sample.

\item In order to get a value of $A_i^{CP}(t_1,t_2)$ as large as possible,
two more predictions have been made with the $K\pi$ invariant mass intervals selected at $0.70~\text{GeV}<\sqrt{s}< 0.75~\text{GeV}$ and $1.40~\text{GeV}<\sqrt{s}< 1.50~\text{GeV}$. It is particularly fascinating to note that the value of this observable in the mass interval $[0.70,0.75]~\text{GeV}$ is as large as the SM prediction for the decay-rate asymmetry~\cite{Grossman:2011zk}.
\end{enumerate}

With the fruitful $\tau$ physics program at Belle II, we hope that the study made in this work can lead to further dedicated measurements of \cp-violating observables in hadronic $\tau$ decays.

\section*{Acknowledgements}
This work is supported by the National Natural Science Foundation of China under Grant Nos.~11675061 and 11775092. X.L. is also supported in part by the Fundamental Research Funds for the Central Universities under Grant No.~CCNU18TS029.

\appendix
\renewcommand{\theequation}{A.\arabic{equation}}
\section*{Appendix: \boldmath The $K\pi$ vector and scalar form factors}
\label{app:ff}
\begin{figure}[ht]
	\centering
	\includegraphics[width=0.45\textwidth]{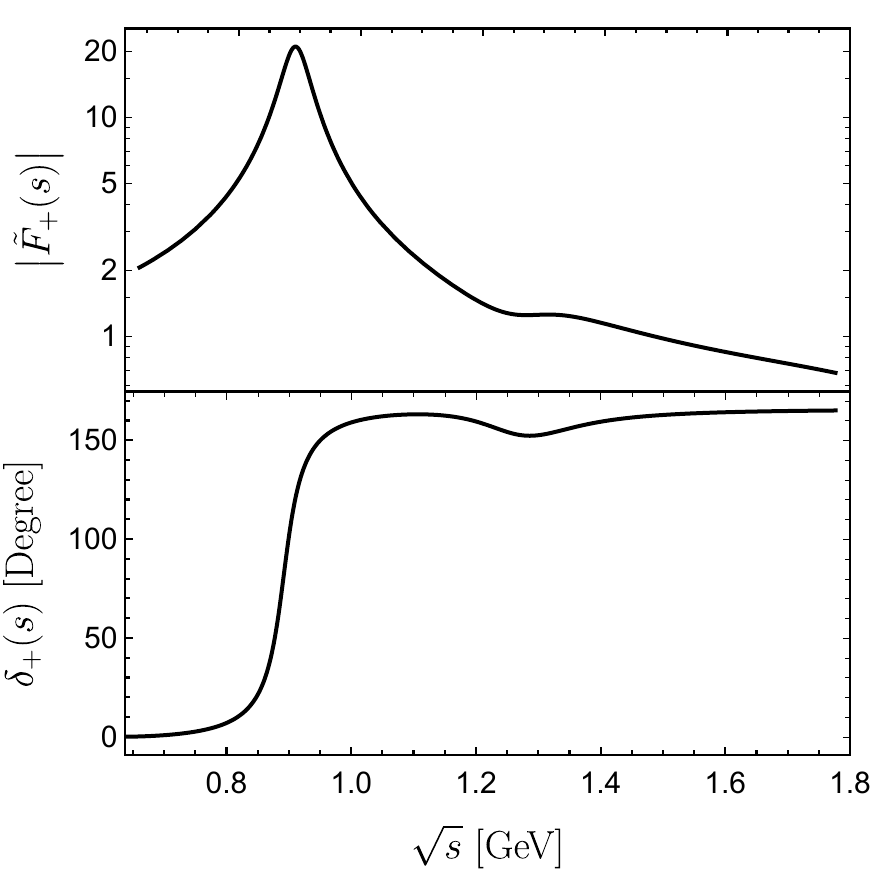}
	\hspace{0.45in}
	\includegraphics[width=0.45\textwidth]{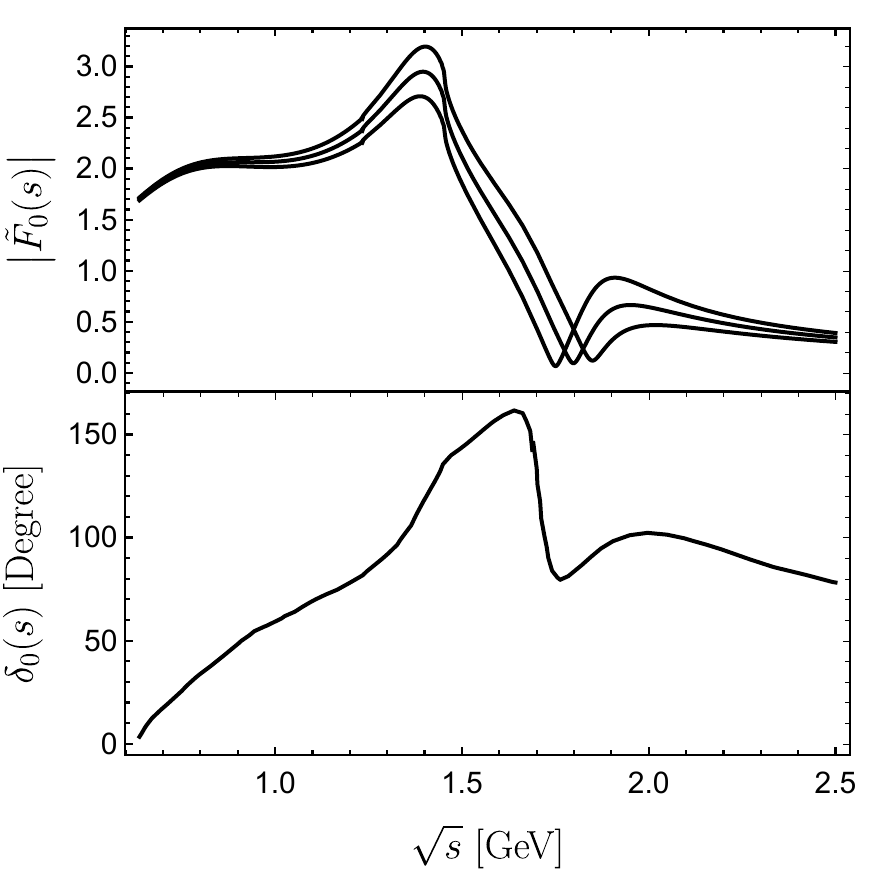}
	\caption{\label{fig:FFs} \small Moduli and phases of the $K\pi$ vector (left) and scalar (right) form factors. The vector form factor is taken from Refs.~\cite{Boito:2008fq,Boito:2010me}, while the scalar form factor is from Ref.~\cite{Jamin:2001zq,Escribano:2014joa}, with the ranges of the modulus obtained by varying the form factor at the Callan-Treiman point~\cite{Jamin:2006tj}.}
\end{figure}

For the normalized $K\pi$ vector form factor, we shall adopt the thrice-subtracted dispersion representation~\cite{Boito:2008fq,Boito:2010me}
\begin{align}\label{eq:vff}
\tilde F_+(s)=\text{exp}\left\lbrace \lambda_+^{\prime}\frac{s}{M_{\pi^-}^2}+\frac{1}{2}(\lambda_+^{\prime\prime}-\lambda_+^{\prime\,2})\frac{s^2}
{M_{\pi^-}^4}+\frac{s^3}{\pi}\int_{s_{K\pi}}^{s_{cut}} ds'\frac{\delta_+(s')}{(s')^3(s'-s-i\epsilon)}\right\rbrace\,,
\end{align}
where one subtraction constant is fixed by the form-factor normalization $F_+(0)=1$, while the other two $\lambda_+^{\prime}$ and $\lambda_+^{\prime\prime}$ describe the slope and curvature of $\tilde F_+(s)$ when performing its Taylor expansion around $s=0$, and hence encode the low-energy behaviour of $\tilde F_+(s)$. To capture our ignorance of the higher-energy part of the dispersion integral, they will be treated as free parameters, and are determined by fitting to the experimental data~\cite{Boito:2008fq,Boito:2010me,Escribano:2014joa}. The form-factor phase $\delta_+(s)$ in Eq.~\eqref{eq:vff} is calculated from the relation
\begin{align}
\tan\delta_+(s)=\frac{\Im m[\tilde f_+(s)]}{\Re e[\tilde f_+(s)]}\,,
\end{align}
where the explicit expression of $\tilde f_+(s)$ has been given by Eq.~(4.1) of Ref.~\cite{Boito:2008fq}, which is derived in the context of chiral perturbation theory with resonances (R$\chi$T)~\cite{Ecker:1988te,Ecker:1989yg}, with both $K^\ast(892)$ and $K^\ast(1410)$ included as explicit degrees of freedom~\cite{Boito:2008fq,Boito:2010me,Jamin:2006tk,Jamin:2008qg}. The cut-off $s_{cut}$ is introduced in Eq.~\eqref{eq:vff} to quantify the suppression of the higher-energy part of the integral, and the stability of the numerical results has been checked by varying $s_{cut}$ in the range $m_\tau<\sqrt{s_{cut}}<\infty$~\cite{Boito:2008fq,Boito:2010me}. Here we shall choose $s_{cut}=4~\text{GeV}^2$, because such a choice is, on the one hand, large enough to not spoil the \textit{a priori} infinite interval of the dispersive integral and, on the other hand, low enough to have a good description of the form-factor phase within the interval considered~\cite{Gonzalez-Solis:2019iod}. Following such a procedure~\cite{Boito:2008fq,Boito:2010me,Gonzalez-Solis:2019iod}, we show in the left panel of Fig.~\ref{fig:FFs} both the modulus and the phase of the normalized form factor $\tilde{F}_+(s)$.

For the $K\pi$ scalar form factor, we shall employ the coupled-channel dispersive representation presented in Ref.~\cite{Jamin:2001zq} and updated later in Refs.~\cite{Jamin:2001zr,Jamin:2004re,Jamin:2006tj}, which is obtained by solving the multi-channel Muskelishivili-Omn\`es problem for three channels (with $1\equiv K\pi$, $2\equiv K\eta$, and $3\equiv K\eta'$). Explicitly, the scalar form factor for the channel $i$ can be written as~\cite{Jamin:2001zq}
\begin{align}\label{eq:sff}
F_0^i(s)=\frac{1}{\pi}\sum\limits_{j=1}^3\int_{s_j}^\infty ds'\frac{\sigma_j(s^\prime)F_0^j(s^\prime)t_0^{i\to j}(s^\prime)^*}{s^\prime-s-i\epsilon}\,,
\end{align}
where $s_j$ and $\sigma_j(s)$ denote respectively the threshold and the two-body phase-space factor for the channel $j$, and $t_0^{i\to j}(s)$ is the partial wave $T$-matrix element for the scattering $i\to j$~\cite{Jamin:2000wn,Jamin:2001zq}. As these form factors are coupled to each other, they can be obtained by solving numerically the coupled dispersion relations arising from Eq.~\eqref{eq:sff}, taking into account the chiral symmetry constraints at low energies as well as the short-distance dynamical QCD constraints at high energies~\cite{Jamin:2000wn,Jamin:2001zq}. Here we shall make use of the numerical results obtained from a combined analysis of the $\tau^-\to K_S\pi^-\nu_\tau$ and $\tau^-\to K^-\eta\nu_\tau$ decays~\cite{Escribano:2014joa}.\footnote{We thank Pablo Roig for providing us with the necessary numerical tables obtained in Ref.~\cite{Escribano:2014joa}.} Again, both the modulus and phase of the reduced form factor $\tilde{F}_0(s)$ are shown in the right panel of Fig.~\ref{fig:FFs}.

\bibliographystyle{JHEP}
\bibliography{reference}

\end{document}